\documentclass[journal]{IEEEtran}

\usepackage{cite}
\usepackage{amsmath,amssymb,amsfonts}
\usepackage{graphicx}
\graphicspath{{figs/}}
\usepackage{textcomp}
\usepackage{booktabs}  
\usepackage{url}
\usepackage{hyperref}
\hypersetup{hidelinks=true}

\def\BibTeX{{\rm B\kern-.05em{\sc i\kern-.025em b}\kern-.08em
    T\kern-.1667em\lower.7ex\hbox{E}\kern-.125emX}}

\begin{document}

\title{Benchmarking the Utility of Privacy-Preserving\\Cox Regression Under Data-Driven Clipping Bounds:\\A Multi-Dataset Simulation Study}

\author{Keita~Fukuyama,
Yukiko~Mori, Tomohiro~Kuroda, and~Hiroaki~Kikuchi
\thanks{Manuscript submitted April 20, 2026.
This work was supported by JST, CREST Grant Number JPMJCR21M1, Japan.}%
\thanks{K. Fukuyama, Y. Mori, and T. Kuroda are with
the Division of Medical Information Technology and Administration Planning, Kyoto University Hospital, Kyoto, Japan (e-mail: fukuyama.keita.45c@st.kyoto-u.ac.jp).}%
\thanks{H. Kikuchi is with
the School of Interdisciplinary Mathematical Sciences, Meiji University, Tokyo, Japan.}}

\markboth{\hskip25pc IEEE JOURNAL OF BIOMEDICAL AND HEALTH INFORMATICS}%
{Fukuyama \MakeLowercase{\textit{et al.}}: Statistical Utility of Differentially Private Cox Regression}

\maketitle

\begin{abstract}

Differential privacy (DP) is a mathematical framework that guarantees individual privacy; however, systematic evaluation of its impact on statistical utility in survival analyses remains limited.
In this study, we systematically evaluated the impact of DP mechanisms (Laplace mechanism and Randomized Response) with data-driven clipping bounds on the Cox proportional hazards model, using 5 clinical datasets ($n = 168$--$6{,}524$), 15 levels of $\varepsilon$ (0.1--1000), and $B = 1{,}000$ Monte Carlo iterations.
The data-driven clipping bounds used here are observed min/max and therefore do not provide formal $\varepsilon$-DP guarantees; the results represent an optimistic lower bound on utility degradation under formal DP.
We compared three types of input perturbations (covariates only, all inputs, and the discrete-time model) with output perturbations (dfbeta-based sensitivity), using loss of significance rate (LSR), C-index, and coefficient bias as metrics.
At standard DP levels ($\varepsilon \leq 1$), approximately 90\% (90--94\%) of the significant covariates lost significance, even in the largest dataset ($n = 6{,}524$), and the predictive performance approached random levels (test C-index $\approx 0.5$) under many conditions.
Among the input perturbation approaches, perturbing only covariates preserved the risk-set structure and achieved the best recovery, whereas output perturbation (dfbeta-based sensitivity) maintained near-baseline performance at $\varepsilon \geq 5$.
At $n \approx 3{,}000$, the significance recovered rapidly at $\varepsilon = 3$--10; however, in practice, $\varepsilon \geq 10$ (for predictive performance) to $\varepsilon \geq 30$--60 (for significance preservation) is required.
In the moderate-to-high $\varepsilon$ range, false-positive rates increased for variables whose baseline $p$-values were near the significance threshold.
\end{abstract}


\section{Introduction}

\IEEEPARstart{D}{ifferential Privacy} (DP)%
\cite{dwork2006calibrating,dwork2014algorithmic} is a mathematical framework that probabilistically limits the influence of any single individual's presence or absence on the output in a database.
As the secondary use of medical data and multi-institutional collaborative research become more popular, balancing patient privacy and statistical utility poses a major challenge.
DP is an emerging standard privacy-protection technique and deployed by Apple Inc., Google, and the U.S.\ Census Bureau.
In clinical trials and epidemiological studies, the Cox proportional-hazards (PH) model \cite{cox1972regression} remains the most widely used survival-analysis method for evaluating treatment effects and prognostic factors. However, applying DP to the Cox model poses several technical challenges.
First, the partial likelihood function relies on the risk set $\mathcal{R}(t) = \{j : T_j \geq t\}$, and adding noise to the survival times $T$ disrupts the ordering structure of the risk set. Second, the privacy budget is diluted by a factor of $q$ ($\varepsilon/q$); therefore, the noise per variable increases in high-dimensional data. Third, medical data typically have $n = 100$--$10{,}000$, which amplifies the relative impact of DP noise.
Despite these challenges, no studies have systematically quantified the extent to which DP degrades the Cox model's statistical utility.

Although previous work at the intersection of DP and survival analysis has addressed output/objective-function perturbation and survival-function estimation (Section~II), a systematic evaluation of \textbf{input perturbation}, a stepwise comparison of perturbation targets, and a quantitative evaluation of utility across dataset sizes remain lacking.
The answer to the question of “can significance be preserved at this $\varepsilon$?” is lacking.

This study aimed to address the following three research questions:
\begin{itemize}
  \item \textbf{RQ1}: At what privacy budget $\varepsilon$ do
    significant covariates in the Cox model lose statistical significance?
    How does dataset size affect this threshold?
  \item \textbf{RQ2}: How does the scope of perturbation targets
    (covariates only / all inputs / discretization) affect utility?
  \item \textbf{RQ3}: At a fixed $\varepsilon$ budget, how does
    statistical utility compare between Local DP (input perturbation)
    and Central DP (output perturbation)?
\end{itemize}
To answer these questions, we conducted simulations using 5 clinical datasets ($n = 168$--$6{,}524$), 15 levels of $\varepsilon$, and $B = 1{,}000$ Monte Carlo iterations.
The primary contributions of this study are as follows: (1) the first systematic evaluation of three input perturbation approaches and output perturbation across 5 datasets $\times$ 15 $\varepsilon$ levels; (2) direct quantification of inferential-utility loss via the loss of significance rate (LSR) under the evaluated DP-style perturbation; (3) elucidation of the scaling effect of significance recovery in the transition zone $\varepsilon = 3$--10; and (4) practical, scope-qualified $\varepsilon$-threshold guidelines for medical researchers.

This study uses data-driven clipping bounds (observed min/max), and formal DP guarantees do not hold.
The results are presented as an \textbf{optimistic lower bound} on utility degradation under minimal-noise conditions.

\section{Related Work}

\subsection{DP and Survival Analysis}

At the intersection of DP and survival analysis, Nguyen and Hui \cite{nguyen2017differentially} proposed DP strategies based on output and objective function perturbations for discrete-time survival regression, achieving coefficient estimation under formal DP guarantees.
Phase~3 (Survival Stacking) of this study is based on the same discrete-time model, but adopts input perturbation (direct noise addition to covariates, exit interval $k^*_i$, and event indicator $\delta_i$), a different approach from the output/objective function perturbation method of Nguyen \& Hui.

Bonomi et al.\ \cite{bonomi2020protecting} proposed a framework for Kaplan--Meier estimation \cite{kaplan1958nonparametric} under DP, addressing privacy protection in survival function estimation. Gondara and Wang \cite{gondara2020differentially} examined DP survival function estimation and confidence interval construction, while De Faveri et al.\ \cite{defaveri2024survival} analyzed the application of a Randomized Response in survival analysis.
In conjunction with federated learning, Rahimian et al.\ \cite{rahimian2022practical} investigated practical challenges in federated DP survival analysis; meanwhile, Hung \& Yu \cite{hung2025federated} theoretically analyzed the minimax optimality of federated DP Cox regression.

\subsection{DP and Regression Analysis}

As general approaches for regression analysis under DP, Chaudhuri et al.\ \cite{chaudhuri2011differentially} proposed output and objective perturbations for empirical risk minimization, while Zhang et al.\ \cite{zhang2012functional} developed a regression analysis using the functional mechanism.
However, since the special structure of the partial likelihood function of the Cox model relies on the risk set $\mathcal{R}(t)$, the direct application of these methods warrants additional theoretical analysis.

\subsection{DP and Hypothesis Testing}

The hypothesis-testing interpretation of DP was first introduced by Wasserman and Zhou \cite{wasserman2010statistical}, who demonstrated that DP mechanisms inherently limit the power of hypothesis tests regarding database membership.
Balle et al.\ \cite{balle2020hypothesis} generalized this result to privacy definitions based on R\'{e}nyi divergence, clarifying the conditions under which the hypothesis-testing interpretation holds.
The loss of significance rate (LSR) used in this study is an empirical measure for testing the general principle that DP noise impairs statistical inference.
These previous studies focused on output/objective function perturbation or survival function estimation; the systematic evaluation of input perturbation, stepwise comparison of perturbation targets, and scaling effects of utility across dataset sizes remain unexplored.
This study fills this knowledge gap by explicitly distinguishing between the trust models of Local DP (input perturbation) and Central DP (output perturbation).

\section{Methods}

\subsection{Fundamentals of Differential Privacy}

A randomized mechanism $\mathcal{M}$ satisfies $\varepsilon$-differential privacy if for any neighboring datasets $D, D'$ (differing by the addition or removal of one record), and any output set $S$, the following holds \cite{dwork2006calibrating}:
\begin{equation}
  \Pr[\mathcal{M}(D) \in S] \leq e^{\varepsilon} \cdot \Pr[\mathcal{M}(D') \in S]
  \label{eq:dp}
\end{equation}

A smaller $\varepsilon$ provides stronger privacy protection, although requires more noise.
This tradeoff between $\varepsilon$ and statistical utility is the central challenge faced by the DP.

A Laplace mechanism was applied to the continuous values.
For the global sensitivity of the function $f$, $\Delta f = \max_{D, D'} \|f(D) - f(D')\|_1$, the noise addition $f(D) + \mathrm{Lap}(\Delta f / \varepsilon)$ satisfies $\varepsilon$-DP.
For discrete values, Randomized Response \cite{warner1965randomized} is used.
For the binary variable $b \in \{0, 1\}$, the true value is retained with probability $e^{\varepsilon}/(1 + e^{\varepsilon})$ and flipped with probability $1/(1 + e^{\varepsilon})$.
For a variable with $k$ categories, this is generalized as follows: The true category is retained with probability $e^{\varepsilon}/(e^{\varepsilon} + k - 1)$, while each category is reported with probability $1/(e^{\varepsilon} + k - 1)$.

When applying DP to $q$ variables, the budget is equally allocated as $\varepsilon/q$ per variable (total $\varepsilon$) by the sequential composition theorem \cite{dwork2014algorithmic}.
Equal allocation is a design choice that assumes no prior information about variable importance, whereas adaptive allocation entails incurring privacy costs for the allocation decision itself and raises issues for post-hoc analysis.

This study evaluates two trust models.
\begin{itemize}
  \item \textbf{Local DP} (input perturbation, Phase~1--3):
    Each individual adds noise to their data before providing it
    to the curator.
    No trust in the curator is required,
    but larger noise is necessary.
  \item \textbf{Central DP} (output perturbation):
    A trusted curator aggregates the raw data
    and adds noise to the statistics.
    Since the influence of one individual is $O(1/n)$,
    structurally less noise is required.
\end{itemize}
The performance difference between the two is consistent with the information-theoretic $\sqrt{n}$ gap \cite{duchi2013local} and does not reflect the superiority of either implementation.

\subsection{Target Datasets}

The five clinical datasets included in the R survival package \cite{therneau2024survival,therneau2000modeling} were used (Table~\ref{tab:baseline}).
The effect of sample size on DP utility was evaluated by spanning sample sizes from small (lung, $n = 168$) to large (flchain,$n = 6{,}524$).
The number of covariates ranges from $q = 7$ to 17, enabling cross-sectional evaluation of the budget dilution effect $\varepsilon/q$.
None of the datasets contained missing values after preprocessing.

\begin{table*}[t]
  \caption{Characteristics of target datasets and baseline Cox regression results.
    C-index is in-sample; Test C-index is the out-of-sample value
    from 70/30 stratified splits ($B = 1{,}000$ iteration mean).
    Sig.~variables is the number of covariates with $p < 0.05$
    at baseline ($\varepsilon = \infty$).}
  \label{tab:baseline}
  \centering
  \footnotesize
  \begin{tabular}{lrrrrrrl}
  \toprule
Dataset & N & Events & q & Event rate & C-index & Test C-index & Sig. variables \\ 
  \midrule
lung & 168 & 121 & 7 & 0.720 & 0.651 & 0.614 & 3/7 (sex, ph.ecog, ph.karno) \\ 
  pbc & 312 & 125 & 17 & 0.401 & 0.857 & 0.831 & 8/17 (age, edema, bili, albumin, copper, ast, protime, stage) \\ 
  colon & 929 & 468 & 11 & 0.504 & 0.671 & 0.659 & 5/11 (nodes, extent, surg, node4, rxLev+5FU) \\ 
  rotterdam & 2982 & 1518 & 9 & 0.509 & 0.679 & 0.675 & 6/9 (age, meno, size>50, size20-50, grade, nodes) \\ 
  flchain & 6524 & 1962 & 7 & 0.301 & 0.790 & 0.791 & 4/7 (age, sexM, sample.yr, lambda) \\ 
   \bottomrule
\end{tabular}

\end{table*}

\subsection{Input Perturbation (Phase~1--3)}

In the input perturbation setting, DP mechanisms are directly applied to each individual's data (Local DP model).
The three Phases are designed with a progressively expanding scope of perturbations.

\subsubsection{Phase~1: Covariate-Only Perturbation}

Only the covariates $\mathbf{X}_i$ are perturbed, whereas the survival time $T_i$ and event indicator $\delta_i$ remain unperturbed.
For the continuous covariates $X_{ij}$, the Laplace mechanism was applied as follows:
\begin{equation}
  X_{ij}^{\mathrm{DP}} = \mathrm{clamp}\!\left(
    X_{ij} + \mathrm{Lap}\!\left(\frac{\mathrm{range}_j}{\varepsilon_j}\right),\;
    \min_j,\; \max_j
  \right)
  \label{eq:laplace_input}
\end{equation}
where $\mathrm{range}_j = \max_j - \min_j$ is the range of covariates $j$ and $\varepsilon_j = \varepsilon / q$ is the budget allocation for variable $j$.
Randomized responses were applied to binary and categorical covariates.
Phase~1 completely preserves the risk-set structure $\mathcal{R}(t) = \{j : T_j \geq t\}$, providing an upper bound (best case) for utility recovery.
However, since $T/\delta$ is provided unperturbed, it constitutes a covariate-level Local DP, not a full Local DP.

\subsubsection{Phase~2: All-Input Perturbation}
In addition to the covariates, Laplace noise was applied to the survival time $T_i$ and Randomized Response to the event indicator $\delta_i$.
Budget allocation was diluted to $\varepsilon/(q+2)$.
The noise in $T$ reverses the ordering of the observation times and induces discontinuous alternation in the risk set membership.
Phase~2 shows the lower bound of the utility degradation when all inputs are protected.

\subsubsection{Phase~3: Discrete-Time Model (Survival Stacking)}
The observation times were partitioned into $K$ data-driven intervals using Sturges' formula, $K = \min(d_{\mathrm{unique}},\; 1 + \lfloor \log_2 N \rfloor)$ ($d_{\mathrm{unique}}$: number of unique event times) ($K = 8$--13 in this study; Table~\ref{tab:sim_params}) \cite{singer1993time,allison1982discrete}.
Each individual's exit interval $k^*_i \in \{1, \ldots, K\}$ (the interval in which observation time $T_i$ belongs) and event indicator $\delta_i$ are independently perturbed.
A Categorical Randomized Response ($K$ categories) is applied to $k^*_i$, and a binary Randomized Response to $\delta_i$.
The expansion based on perturbed $(k^{*\prime}_i, \delta'_i)$ (row generation from interval $1$ to $k^{*\prime}_i$) does not incur additional privacy costs according to the post-processing theorem.
The budget allocation is identical to Phase~2 at $\varepsilon/(q+2)$, enabling a direct comparison between continuous perturbation (Phase~2) and discretization (Phase~3).
Discrete-time hazards were estimated using logistic regression.
This approach represents an input-perturbation variant of the discrete-time model, for which Nguyen \& Hui \cite{nguyen2017differentially} achieved a formal DP via an output/objective-function perturbation.

Notably, since Phase~3 uses a discrete-time generalized linear model (GLM) rather than Cox PH, three variables (flchain kappa, lung ph.karno, flchain sample.yr) yield different significance determinations even at $\varepsilon = \infty$ (no DP).
Because the baseline significance/nonsignificance of these variables changes owing to the model change, the LSR/FPR attributable to DP cannot be discussed.
Therefore, these were excluded from the phase ~3 LSR/FPR calculations.

\subsection{Output Perturbation (Central DP)}

In the output perturbation, Cox regression was performed on the unperturbed data, and Laplace noise was added to the estimated regression coefficients $\hat{\boldsymbol{\beta}}$.
The dfbeta residuals of the Cox regression (a first-order approximation of the leave-one-out influence function) \cite{therneau2000modeling} were used as sensitivity:
\begin{equation}
  \Delta\hat{\beta}_j = \max_i |\mathrm{dfbeta}_{i,j}|
  \label{eq:dfbeta_sens}
\end{equation}
Similar to the data-driven clipping bounds (observed min/max) for the input perturbation, the dfbeta sensitivity is also data-dependent, and formal DP guarantees do not hold; however, the aim is utility evaluation under minimum-noise conditions.

\subsection{Evaluation Metrics}

\subsubsection{Loss of Significance Rate (LSR)}
The proportion of covariates that were significant ($p < 0.05$) before DP application and became non-significant after DP application across $B$ simulations.
\begin{equation}
  \mathrm{LSR}_j = \frac{1}{B}\sum_{b=1}^{B}
    \mathbb{1}\!\left[p_j^{\mathrm{clean}} < 0.05 \;\wedge\;
    p_{j,\mathrm{DP}}^{(b)} \geq 0.05\right]
  \label{eq:lsr}
\end{equation}
LSR is an empirical observation in regression coefficient significance testing of the general principle that DP noise degrades statistical inference, consistent with the privacy--power trade-off of Balle et al.\ \cite{balle2020hypothesis}.
This study reported the mean LSR across all significant covariates within each dataset.

\subsubsection{C-index}
The Concordance index \cite{harrell1982evaluating} measures the model's predictive ranking accuracy.
$C = 0.5$ indicates random prediction, and $C = 1.0$ indicates perfect discrimination.
For out-of-sample evaluation, the test C-index was computed using 70/30 stratified splits with $B = 1{,}000$ Monte Carlo cross-validation iterations.
Out-of-sample evaluation complemented the in-sample C-index, highlighting the impact of DP noise on generalization performance.
The C-index degradation $\Delta C = \bar{C}_{\mathrm{test}}(\varepsilon = \infty) - \bar{C}_{\mathrm{test}}(\varepsilon)$ has also been reported.

\subsubsection{False Positive Rate (FPR)}
The proportion of covariates that were not significant ($p \geq 0.05$) before DP application became significant after DP application.
\begin{equation}
  \mathrm{FPR}_j = \frac{1}{B}\sum_{b=1}^{B}
    \mathbb{1}\!\left[p_j^{\mathrm{clean}} \geq 0.05 \;\wedge\;
    p_{j,\mathrm{DP}}^{(b)} < 0.05\right]
  \label{eq:fpr}
\end{equation}
While LSR measures the loss of significance of significant covariates, FPR is a complementary metric that measures the false positives of non-significant covariates.

\subsubsection{Relative Bias of Hazard Ratios}
The hazard ratio (HR) is the primary measure of effect size reported in clinical research, and bias from DP noise directly impairs the clinical interpretation of treatment effects and risk factors.
The signed relative bias is computed as $\mathrm{Bias}_j = B^{-1}\sum_{b}(\mathrm{HR}_{j,\mathrm{DP}}^{(b)} - \mathrm{HR}_j^{\mathrm{clean}}) / \mathrm{HR}_j^{\mathrm{clean}}$.

\subsection{Simulation Design}

Table~\ref{tab:sim_params} lists the simulation parameters.
For each condition (five datasets $\times$ 15 $\varepsilon$ $\times$ four methods), $B = 1{,}000$ independent simulations were executed.
$\varepsilon$ spans 15 levels from 0.1 to 1000, and $\infty$ (Table~\ref{tab:sim_params}); $\varepsilon = 3$ and $\varepsilon = 7$ were added for high-resolution analysis of the transition zone.
Reproducibility was ensured using seed values (see the GitHub repository for implementation details).

\begin{table}[t]
  \caption{Simulation parameters}
  \label{tab:sim_params}
  \centering
  \footnotesize
  \begin{tabular}{lp{0.55\columnwidth}}
  \toprule
Parameter & Value \\
  \midrule
Iterations ($B$) & 1,000 \\
  $\varepsilon$ values & 0.1, 0.5, 1.0, 2.0, 3.0, 5.0, 7.0, 10, 15, 30, 60, 100, 250, 1000, $\infty$ (15 values) \\
  Datasets & lung (168), pbc (312), colon (929), rotterdam (2982), flchain (6524) \\
  Random seed & 42 (reproducible) \\
  Train/test split & 70/30 stratified by event status \\
  Phase 3 $K$ (Sturges) & lung=8, pbc=9, colon=10, rotterdam=12, flchain=13 \\
  DP mechanisms & Laplace mechanism; randomized response \\
  Clipping bounds & Data-driven (observed min/max) \\
   \bottomrule
\end{tabular}

\end{table}

\subsection{Limitations of Data-Driven Bounds}

In this study, the clipping bounds were determined from the observed min/max of the data, which constitutes privacy leakage; therefore, formal DP guarantees do not hold.
Formal DP requires either (a)~data-independent bounds (based on domain knowledge) or (b)~ a privacy budget for bound estimation.
In either case, the noise would be larger than that in our setting. Thus, the results of this study constitute an \textbf{optimistic lower bound} for utility degradation under formal DP.
Similarly, the dfbeta sensitivity of the output perturbation is data-driven, and formal DP guarantees do not hold true.

Relatedly, continuous covariates were retained on their original scale rather than z-standardized, because the per-variable sensitivity $\Delta_j = U_j - L_j$ already scales the Laplace noise to each covariate's observed range; Wald-test-based significance evaluation is scale-invariant, so standardization would not change the reported LSR/FPR results.
Retaining the original scale also preserves the clinical interpretability of hazard-ratio estimates in a deployment scenario.
Implementation-level details (missing-value handling, categorical encoding, tied-event handling, train/test split, and Phase~3 time-grid construction) are documented in the accompanying GitHub repository to keep the Methods focused on the DP mechanism.

\section{Results}

\subsection{Baseline}

Table~\ref{tab:baseline} lists the baseline characteristics of five datasets.
The sample sizes range from $n = 168$ to $6{,}524$ (approximately 40-fold), covering $q = 7$--17 covariates.
The test C-indices were close to the in-sample values (0.614--0.831), confirming the stability of the baseline predictive performance.

\subsection{Loss of Significance Rate (LSR)}

\begin{figure*}[t]
  \centering
  \includegraphics[width=\textwidth]{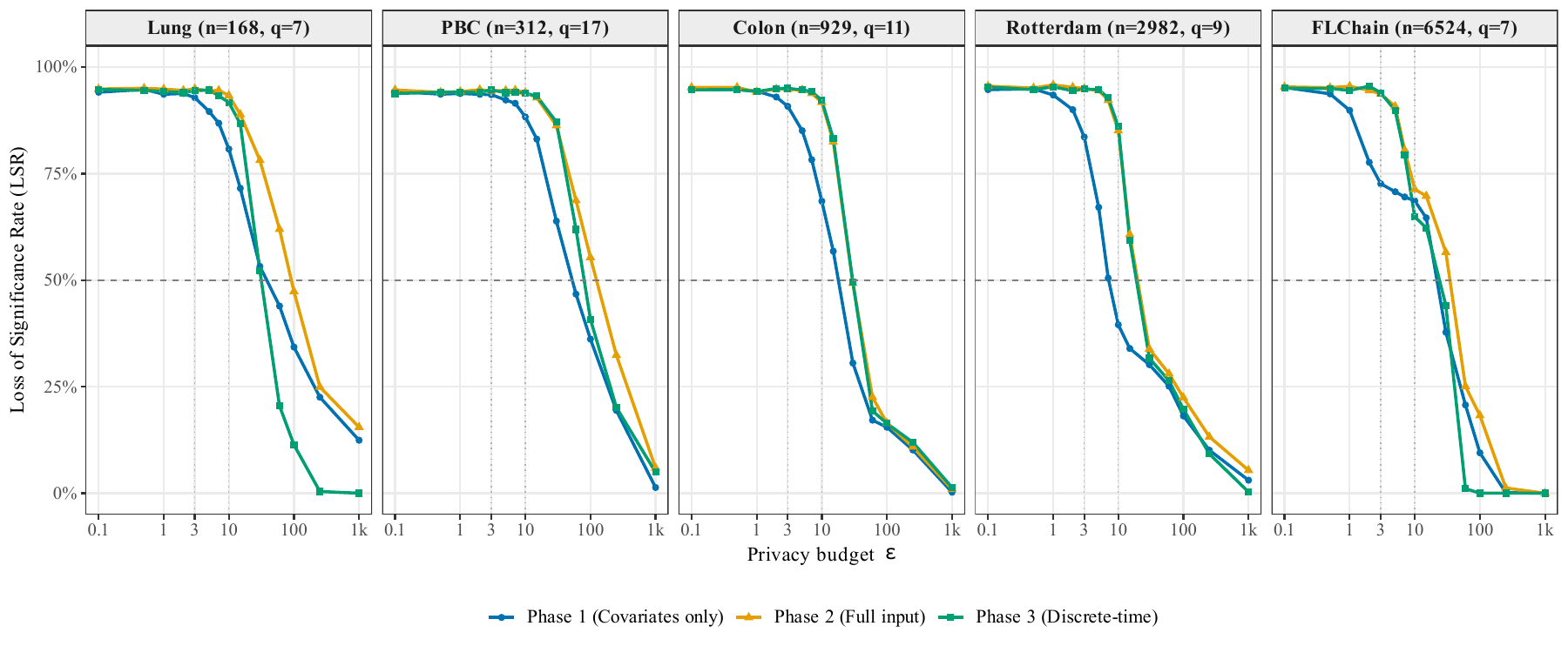}
  \caption{$\varepsilon$-dependence of the loss of significance rate (LSR).
    Three input perturbation approaches (Phase~1--3)
    compared across five datasets.
    Dashed lines indicate $\varepsilon = 3$ and $\varepsilon = 10$.
    Phase~1 (covariates only) achieves the best recovery
    across all $\varepsilon$ ranges;
    Phase~2 (all inputs) is inferior due to risk-set destruction;
    Phase~3 (discrete-time) shows no improvement
    from discretization under the same budget structure.}
  \label{fig:lsr}
\end{figure*}

Fig.~\ref{fig:lsr} shows the $\varepsilon$-dependence of LSR across all datasets for Phases~1--3.
Supplementary Table~S1 provides full numerical values.

\textbf{At standard DP levels ($\varepsilon \leq 1$), the evaluated input-perturbation schemes lose significance across all datasets.} The LSR at $\varepsilon = 1$ for Phase~1 ranged from 0.899 to 0.944 across all datasets, and even the largest dataset, flchain ($n = 6{,}524$), had LSR $= 0.899$ (Supplementary Table~S1).
At $\varepsilon \leq 1$, approximately 90\% (89--94\%) of the significant covariates lost their significance, regardless of the dataset size.

\textbf{Significance recovery depends strongly on $\varepsilon$ and dataset size.} At $\varepsilon = 10$ for Phase~1, rotterdam ($n = 2{,}982$, LSR $= 0.395$) showed the greatest recovery, whereas pbc ($q = 17$, LSR $= 0.883$) showed the slowest recovery (Supplementary Table~S1).
This difference is attributed to the combination of $n$ magnitude and budget dilution $\varepsilon/q$.
The variable-level LSR recovery speed depended on the effect size and sensitivity (Supplementary Fig.~S1).

\textbf{Phase~2 is consistently inferior to Phase~1.} In Phase~2, LSR recovery was markedly delayed due to the destruction of the risk-set structure, remaining at LSR $> 0.90$ for $\varepsilon \leq 5$.
Phase~3 (discrete-time model) yielded comparable results under the same $\varepsilon/(q+2)$ budget structure as in Phase~2, and discretization itself did not contribute to utility improvement.
The convergence failure rate was 0

\subsection{Out-of-Sample Predictive Performance}

\begin{figure*}[t]
  \centering
  \includegraphics[width=\textwidth]{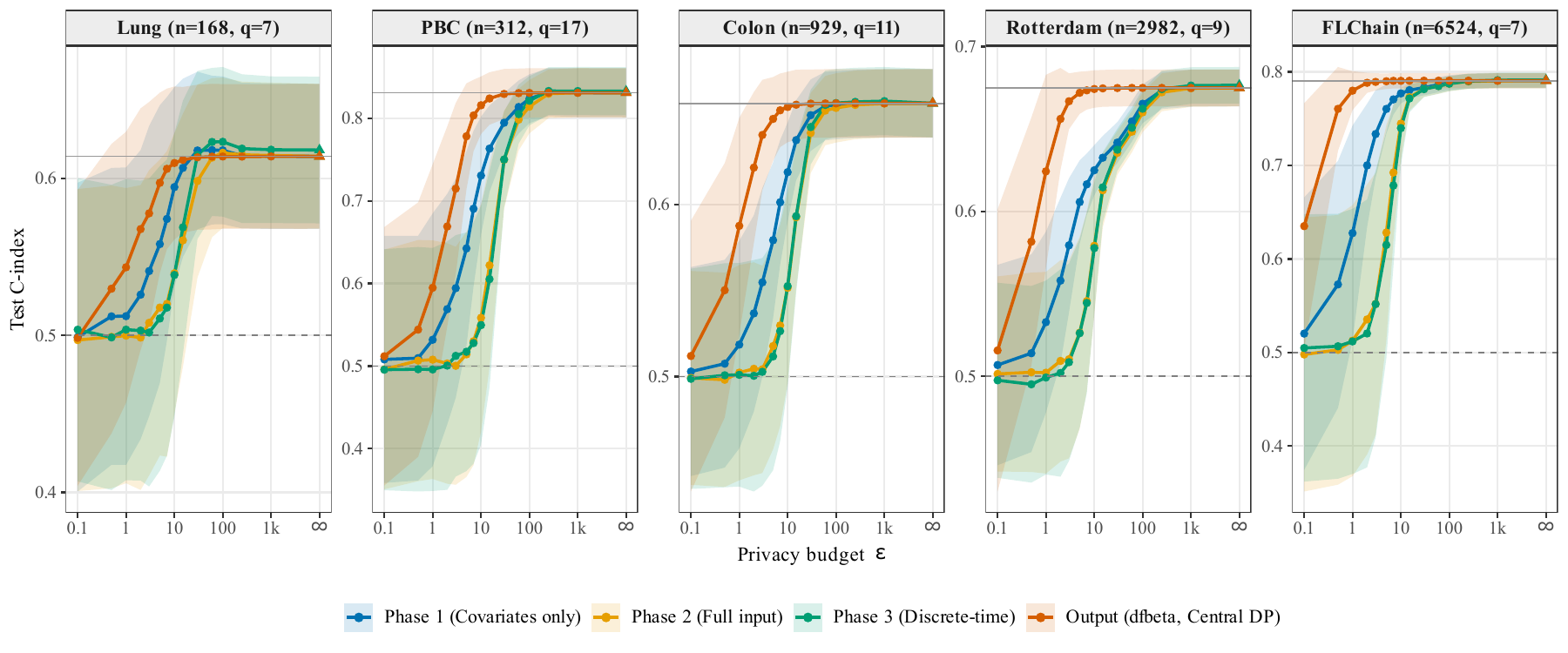}
  \caption{$\varepsilon$-dependence of the test C-index.
    Four methods (Phase~1--3, output perturbation dfbeta)
    compared across five datasets.
    Ribbons indicate $\pm 1$ standard deviation (SD)
    ($B = 1{,}000$ train/test splits).
    Triangle markers at the right end indicate
    the baseline under the non-DP condition ($\varepsilon = \infty$).
    Gray solid lines indicate the baseline C-index; dashed lines indicate random prediction (0.5).}
  \label{fig:cindex}
\end{figure*}

Fig.~\ref{fig:cindex} shows the $\varepsilon$-dependence of the test C-index for Phases~1--3 and the output perturbation (dfbeta).
Supplementary Table~S2 lists the entire numerical values.

\textbf{At $\varepsilon \leq 1$, the C-index collapses to the random prediction level.} The test C-index for Phase~1 degraded to 0.512--0.628 at $\varepsilon = 1$, and Phase~2 saturated at the C-index $\approx 0.5$ for $\varepsilon \leq 1$ (Supplementary Table~S2).
The latter is direct evidence of risk-set destruction.

\textbf{Phase~1 achieves the best utility recovery.} The C-index of Phase~1 at $\varepsilon = 10$ was 0.594--0.777, recovering to more than 90\% of the baseline for large datasets (rotterdam and flchain).
Although DP noise can function as a regularization, the overfitting gap (Train C-index $-$ Test C-index) $\varepsilon$-dependence varies by dataset (Supplementary Fig.~S2).

\textbf{Output perturbation (dfbeta) exhibits the structural characteristics of Central DP.} The dfbeta-based output perturbation maintained a near-baseline performance at $\varepsilon \geq 5$ and outperformed Phase~1 across all $\varepsilon$ ranges in four of the five datasets (Supplementary Table~S2).
Only for the lungs did it become comparable to phase ~1 at $\varepsilon \geq 30$, with a difference of 0.005.

\subsection{False Positive Rate (FPR)}

\begin{figure*}[t]
  \centering
  \includegraphics[width=\textwidth]{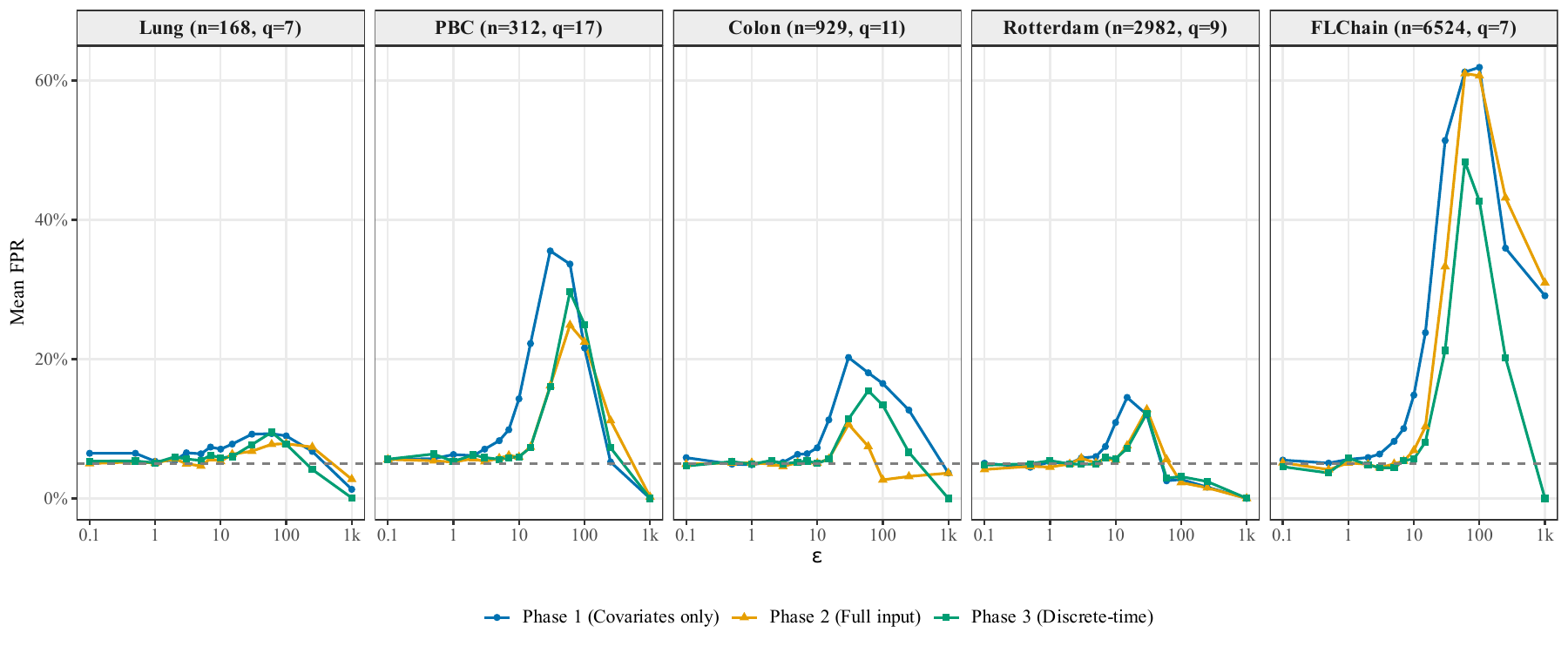}
  \caption{$\varepsilon$-dependence of the mean false positive rate (FPR)
    for three input perturbation methods (Phase~1--3)
    across five datasets.
    The dashed line indicates the nominal $\alpha = 0.05$.
    At low $\varepsilon$, FPR $\approx \alpha$,
    but it increases at moderate-to-high $\varepsilon$ (10--100)
    and converges to 0 as $\varepsilon \to \infty$.}
  \label{fig:fpr}
\end{figure*}

\begin{figure*}[t]
  \centering
  \includegraphics[width=\textwidth]{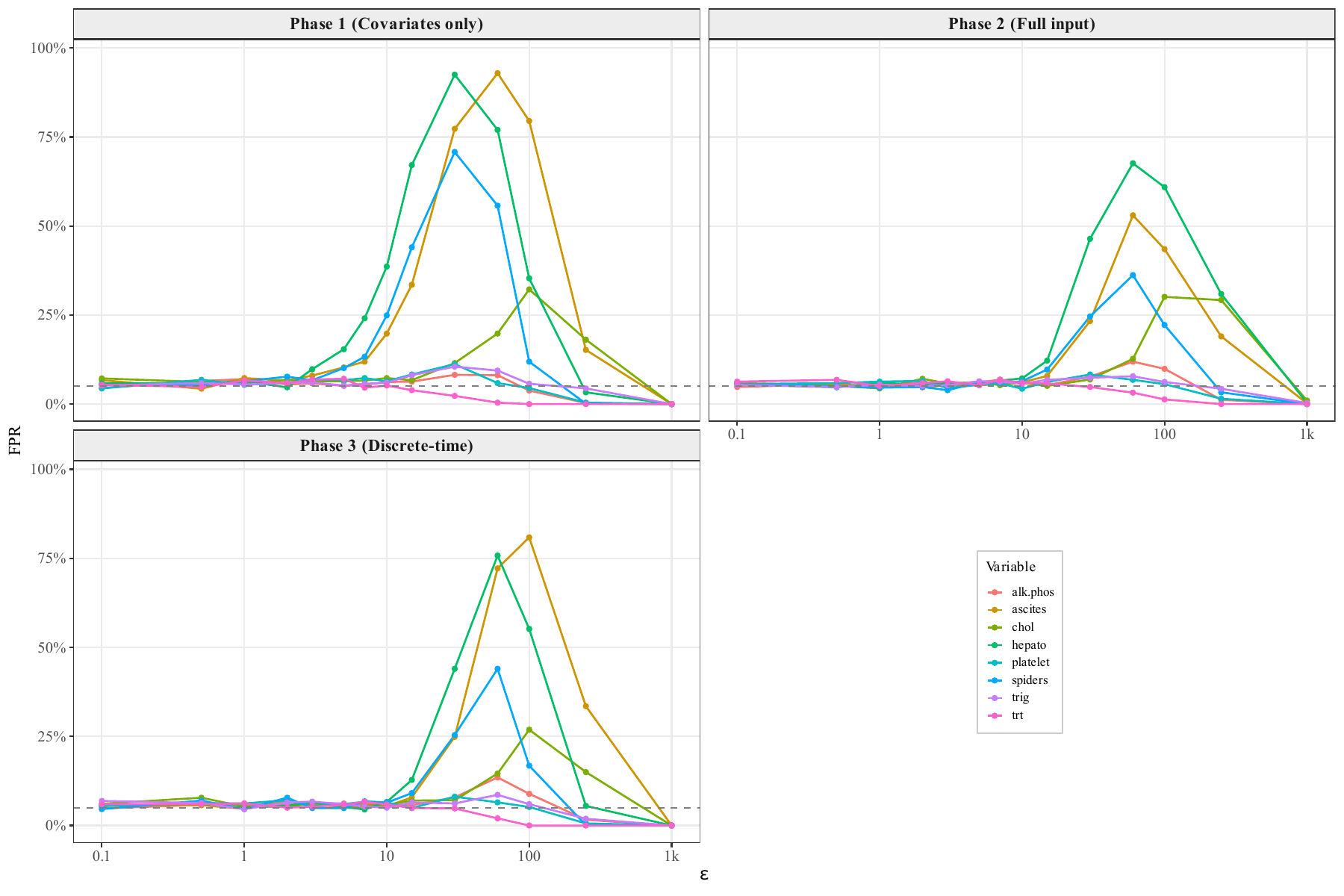}
  \caption{Variable-level FPR for the PBC dataset ($q = 17$)
    by Phase.
    The $y$-axis range is unified across all panels.
    The dashed line indicates the nominal $\alpha = 0.05$ level.
    FPR peaks in the intermediate $\varepsilon$ range (10--100)
    for variables whose baseline $p$-values are near
    the significance threshold,
    and converges to 0 as $\varepsilon \to \infty$.}
  \label{fig:fpr-var}
\end{figure*}

FPR, at which the proportion of covariates that were non-significant at baseline but became significant following DP application, was analyzed for input perturbations (Phases~1--3).
FPR exhibits a three-stage pattern with respect to $\varepsilon$ (Fig.~\ref{fig:fpr}, Fig.~\ref{fig:fpr-var}; Supplementary Table~S3). (1)~At a low $\varepsilon$ ($\leq 1$), FPR $\approx \alpha = 0.05$ remains at the nominal level. (2)~At moderate to high $\varepsilon$ (10--100), FPR increased notably for variables whose baseline $p$-values were close to $\alpha$.
Representative examples include flchain kappa (peak FPR $\approx 100$\%, $\varepsilon = 100$) and colon differences (peak FPR $\approx 93$\%, $\varepsilon = 100$). (3)~As $\varepsilon \to \infty$, FPR $\to 0$ converges to the non-DP condition.
This result indicates that even in the high $\varepsilon$ range (10--100), where significance recovers, the false-positive risk simultaneously increases.
Suppressing FPR to the nominal level requires an even larger $\varepsilon$; caution is warranted for variables whose baseline $p$-values are near the $\alpha$ boundary.

\subsection{Scaling of Dataset Size}

Table~\ref{tab:transition} lists the Phase~1 LSR progression in the transition zone, $\varepsilon = 3$--10.

\textbf{At $\varepsilon \leq 1$, the dataset size effect is nullified.} Since the noise scale of the input perturbation $\mathrm{range}_j / \varepsilon_j$ does not depend on $n$, all five datasets have LSR $> 0.90$ and a C-index $\approx 0.5$ at $\varepsilon \leq 1$.
The effect of $n$ manifests only as the standard error of the estimator ($\propto 1/\sqrt{n}$).

\textbf{$\varepsilon = 3$--10 constitutes the transition zone.} As shown in Table~\ref{tab:transition}, rotterdam ($n = 2{,}982$) substantially declines in LSR from 0.836 to 0.395, whereas the lung ($n = 168$) from 0.928 to 0.808.
Flchain ($n = 6{,}524$) does not recover, whereas the behavior rotterdam observed is attributed to the presence of large-range covariates and a high proportion of non-significant variables. Thus, utility depends on the combined effects of $n$, $\varepsilon$, $q$, and the covariate range.

\begin{table}[t]
  \caption{Phase~1 LSR progression in the transition zone
    ($\varepsilon = 3$--10).}
  \label{tab:transition}
  \centering
  \footnotesize
  \begin{tabular}{lrrrr}
  \toprule
Dataset & $\varepsilon = 3$ & $\varepsilon = 5$ & $\varepsilon = 7$ & $\varepsilon = 10$ \\ 
  \midrule
lung & 0.928 & 0.896 & 0.868 & 0.808 \\ 
  pbc & 0.935 & 0.923 & 0.915 & 0.883 \\ 
  colon & 0.908 & 0.851 & 0.783 & 0.685 \\ 
  rotterdam & 0.836 & 0.671 & 0.505 & 0.395 \\ 
  flchain & 0.726 & 0.708 & 0.695 & 0.686 \\ 
   \bottomrule
\end{tabular}

\end{table}

\subsection{Regression Coefficient Bias}

\begin{figure*}[t]
  \centering
  \includegraphics[width=\textwidth]{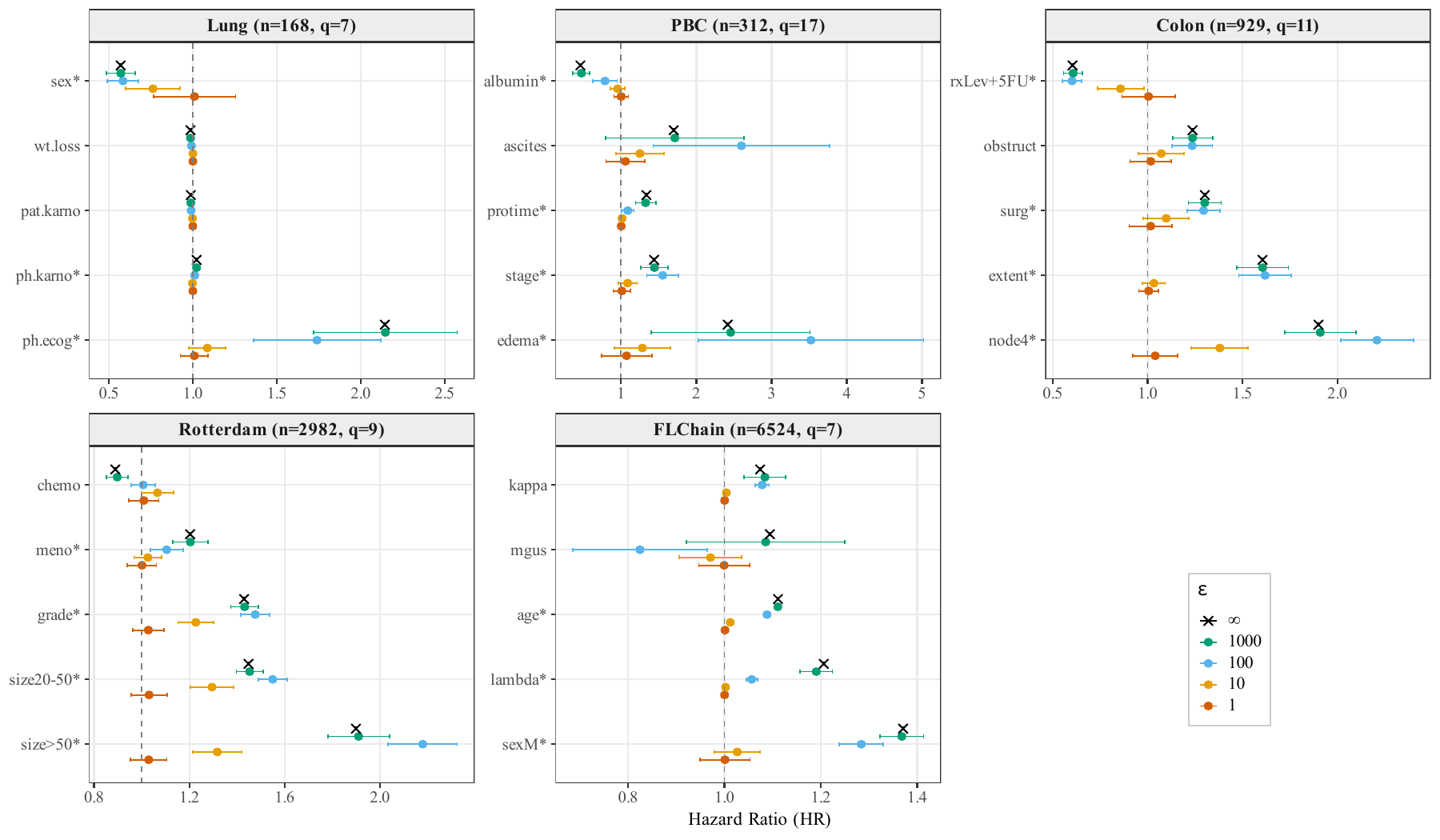}
  \caption{Hazard ratio (HR) distributions for the top five covariates
    in each dataset under Phase~1 (covariate-only perturbation).
    Dots indicate the mean HR across $B = 1{,}000$ iterations;
    error bars indicate $\pm 1$ SD.
    $\times$ marks indicate the baseline HR
    under the non-DP condition ($\varepsilon = \infty$).
    The dashed line indicates HR $= 1.0$ (no effect).
    At low $\varepsilon$, all HRs shrink toward 1.0,
    and effect sizes vanish.
    Variables are ordered by descending baseline regression coefficients.}
  \label{fig:hr_dist}
\end{figure*}

\begin{figure*}[t]
  \centering
  \includegraphics[width=\textwidth]{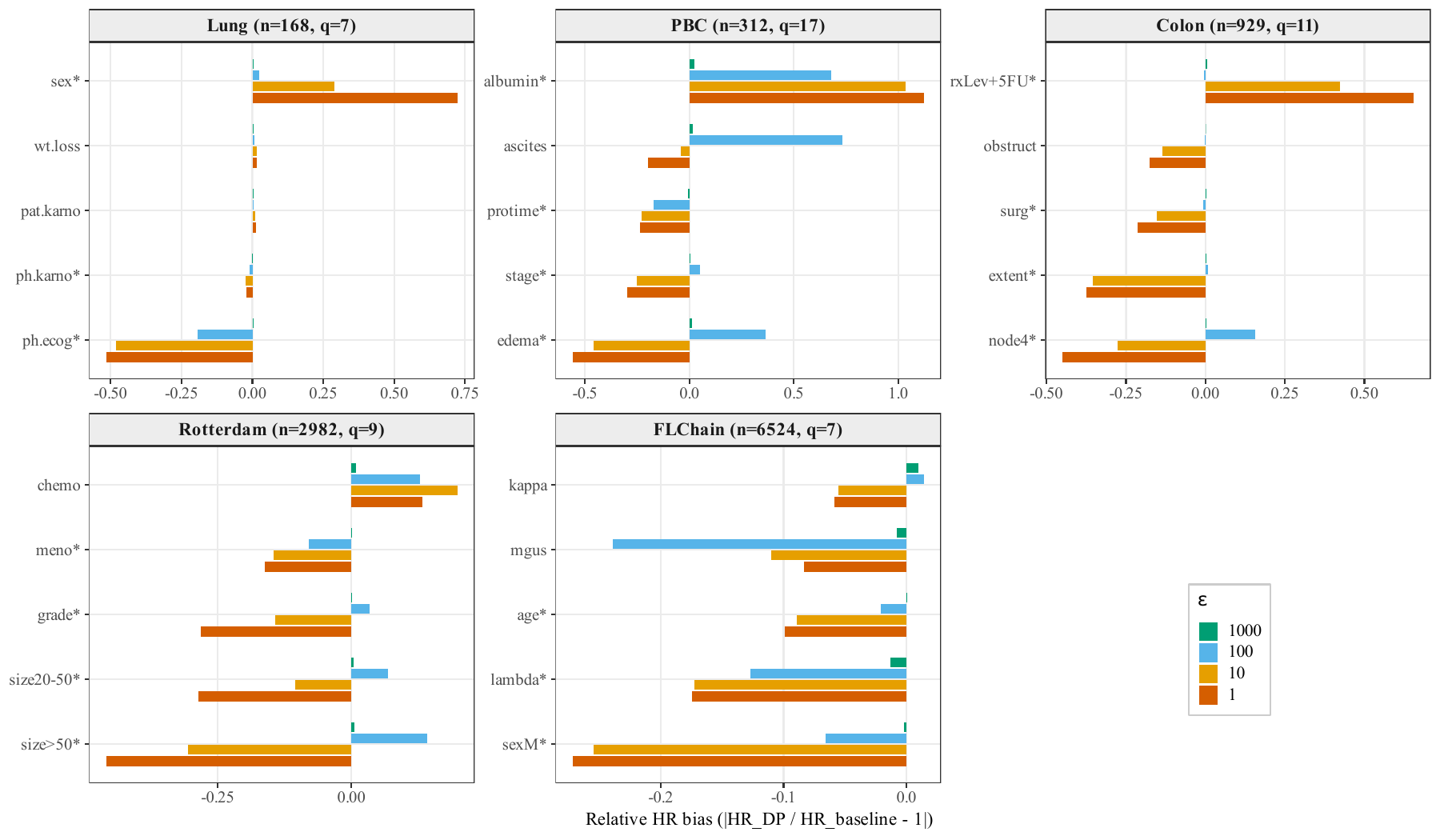}
  \caption{HR relative bias for the top five covariates under Phase~1
    ($|\mathrm{HR_{DP}} / \mathrm{HR_{baseline}} - 1|$).
    Compared at four levels: $\varepsilon \in \{1, 10, 100, 1000\}$.
    $*$ indicates variables significant at baseline ($p < 0.05$).
    Variables are listed by descending baseline regression coefficients
    (same order as Fig.~\ref{fig:hr_dist}).
    At lower $\varepsilon$, HRs shrink toward 1.0,
    demonstrating that DP noise attenuates effect sizes.}
  \label{fig:hr_bias}
\end{figure*}

Fig.~\ref{fig:hr_dist} shows the hazard ratio (HR) distributions of the top five covariates across the five datasets.
Fig.~\ref{fig:hr_bias} presents the HR relative biases for the same variables.

At $\varepsilon = 1$, HRs degenerate toward 1.0 ($\times$ marks = comparison with the baseline HR), making it impossible to even discern the effects' direction (Fig.~\ref{fig:hr_dist}).
At $\varepsilon \geq 100$, HRs approach the baseline, although the bias remains significant at $\varepsilon \leq 5$ even for large datasets, suggesting that increasing $n$ alone cannot compensate for the noise magnitude.

\subsection{Practical Thresholds}

Table~\ref{tab:threshold} lists the practical $\varepsilon$-thresholds for each dataset $\times$ method.
$\varepsilon^*_{\Delta C \leq 0.05}$ is the minimum $\varepsilon$ at which test C-index degradation is within 0.05; $\varepsilon^*_{\mathrm{LSR} \leq 0.50}$ is the minimum $\varepsilon$ at which LSR falls to 50\% or below; $\varepsilon^*_{\mathrm{FPR} \leq 0.10}$ is the minimum $\varepsilon$ at which the mean FPR permanently falls below 10\% (the right tail of the FPR peak).

The $\varepsilon^*_{\Delta C \leq 0.05}$ for Phase~1 ranged from 5 to 30, with larger datasets achieving a practical predictive performance at a lower $\varepsilon$ (Table~\ref{tab:threshold}).
By contrast, significance preservation requires $\varepsilon \geq 10$--60 ($\mathrm{LSR} \leq 0.50$).
FPR exhibited a bell-shaped pattern (Fig.~\ref{fig:fpr}), maintaining the nominal $\alpha$ level at low $\varepsilon$, peaking in the intermediate $\varepsilon$ range (10--100), and converging to zero as $\varepsilon \to \infty$.
$\varepsilon^*_{\mathrm{FPR}}$ is defined as the minimum $\varepsilon$ after which FPR permanently returns to below 10\%.
The $\varepsilon^*_{\mathrm{FPR}}$ for Phase~1 ranged widely from 60 (rotterdam) to $> 1000$ (flchain), with only the lungs having FPR below 10\% across all ranges.
Notably, the $\varepsilon$ range satisfying $\varepsilon^*_{\Delta C}$ or $\varepsilon^*_{\mathrm{LSR}}$ overlaps with the FPR danger zone, indicating that caution is needed when interpreting ``recovered'' significance after DP application at face value.

\begin{table*}[t]
  \caption{Practical $\varepsilon$ thresholds.
    $\varepsilon^*_{\Delta C}$: minimum $\varepsilon$
    for $\Delta C \leq 0.05$ (C-index degradation threshold);
    $\varepsilon^*_{\mathrm{LSR}}$: minimum $\varepsilon$
    for LSR $\leq 0.50$ (half-recovery of significance);
    $\varepsilon^*_{\mathrm{FPR}}$: minimum $\varepsilon$
    at which mean FPR permanently falls below $0.10$
    (right tail of FPR peak).
    --- indicates FPR $< 0.10$ at all $\varepsilon$ (lung)
    or not applicable (output perturbation).}
  \label{tab:threshold}
  \centering
  \footnotesize
  \begin{tabular}{llllll}
  \toprule
Dataset & Phase & $\varepsilon^*_{\Delta C \leq 0.05}$ & $\varepsilon^*_{\text{LSR} \leq 0.5}$ & $\varepsilon^*_{\text{LSR} \leq 0.1}$ & $\varepsilon^*_{\text{FPR} \leq 0.10}$ \\ 
  \midrule
lung & Output & 2 & $>$ 1000 & $>$ 1000 & --- \\ 
  lung & Phase 1 & 7 & 60 & $>$ 1000 & --- \\ 
  lung & Phase 2 & 30 & 100 & $>$ 1000 & --- \\ 
  lung & Phase 3 & 15 & 60 & 250 & --- \\ 
  pbc & Output & 7 & $>$ 1000 & $>$ 1000 & --- \\ 
  pbc & Phase 1 & 30 & 60 & 1000 & 250 \\ 
  pbc & Phase 2 & 60 & 250 & 1000 & 1000 \\ 
  pbc & Phase 3 & 60 & 100 & 1000 & 250 \\ 
  colon & Output & 2 & $>$ 1000 & $>$ 1000 & --- \\ 
  colon & Phase 1 & 10 & 30 & 1000 & 1000 \\ 
  colon & Phase 2 & 30 & 30 & 1000 & 60 \\ 
  colon & Phase 3 & 30 & 30 & 1000 & 250 \\ 
  rotterdam & Output & 2 & $>$ 1000 & $>$ 1000 & --- \\ 
  rotterdam & Phase 1 & 10 & 10 & 1000 & 60 \\ 
  rotterdam & Phase 2 & 30 & 30 & 1000 & 60 \\ 
  rotterdam & Phase 3 & 30 & 30 & 250 & 60 \\ 
  flchain & Output & 0.5 & $>$ 1000 & $>$ 1000 & --- \\ 
  flchain & Phase 1 & 5 & 30 & 100 & $>$ 1000 \\ 
  flchain & Phase 2 & 10 & 60 & 250 & $>$ 1000 \\ 
  flchain & Phase 3 & 15 & 30 & 60 & 1000 \\ 
   \bottomrule
\end{tabular}

\end{table*}

\section{Discussion}

\subsection{$\varepsilon$ Threshold for Significance Collapse (RQ1)}

At standard DP levels ($\varepsilon \leq 1$), the DP-style perturbation schemes evaluated here essentially eliminate meaningful inference on Cox regression coefficients, regardless of dataset size (Section~IV-B).
This result indicates that the noise scale of 
input perturbation $\mathrm{range}_j/\varepsilon_j$ does not depend on $n$.
At $\varepsilon \leq 1$, the noise completely overwhelms the signal, and the effect of $n$ manifests only through the standard error of the estimator ($\propto 1/\sqrt{n}$), which alone cannot compensate for the noise magnitude.

In the transition zone, $\varepsilon = 3$--10, the signal-to-noise ratio begins to allow room for improvement and the effect of $n$ is maximized.
The difference in the recovery speed between rotterdam and the lung (Table~\ref{tab:transition}) reflects this structural asymmetry.

The applicability of DP should be judged after quantifying the specific trade-off between privacy risk and utility loss. The threshold table of this study (Table~\ref{tab:threshold}) provides the basis for such judgment.

\subsection{Vulnerability of Risk Set Structure (RQ2)}

The C-index saturation of Phase~2 (Section~IV-C) demonstrates that noise on $T$ irreversibly disrupts the ordering structure of risk sets.
For the Cox model, whose partial likelihood depends on the risk set $\mathcal{R}(t)$, order reversal in $T$ means a fundamental loss of discriminative ability.
Phase~3 theoretically circumvents the risk-set problem and achieves the same $\varepsilon/(q+2)$ budget structure as Phase~2.
However, discretization under the same budget structure did not contribute to utility improvement, and the high flip rate of categorical Randomized Response constrained performance in the low-$\varepsilon$ range.
When applying DP to Cox regression, avoiding perturbation of $T/\delta$ should therefore be the fundamental design principle.

\subsection{Predictive Utility and Inferential Utility Recover Differently}

A recurring pattern across methods, and particularly for output perturbation, is that \textbf{predictive discrimination (C-index) and inferential significance (LSR) do not recover at the same $\varepsilon$.}
Output perturbation maintains near-baseline test C-index at $\varepsilon \geq 5$ (Supplementary Table~S2), yet its $\varepsilon^*_{\mathrm{LSR} \leq 0.5}$ exceeds 1000 for every dataset (Table~\ref{tab:threshold}); in other words, a model that still ranks high-risk and low-risk patients well can simultaneously fail to declare any individual covariate statistically significant.
This dissociation is not a paradox: dfbeta-based noise is allocated per coefficient and dominates the $\beta_j / \mathrm{SE}(\beta_j)$ ratio used in Wald tests, while discrimination depends on the aggregate linear predictor, which tolerates uncorrelated per-coefficient noise far better.
For clinical deployment this distinction matters: a DP-released Cox model may be safely usable as a risk-stratification tool long before its individual coefficients can be interpreted as effect-size estimates, and LSR-based $\varepsilon$ recommendations should not be mechanically applied to predictive-modeling use cases (or vice versa).
Reporting LSR, C-index, and HR relative bias together---as done throughout Section~IV---is therefore more informative than reporting any one metric in isolation.

\subsection{Local DP vs Central DP (RQ3)}

The output perturbation (dfbeta) outperformed Phase~1 under nearly all conditions (Fig.~\ref{fig:cindex}), with the difference being particularly pronounced for datasets with large $q$ (pbc, $q = 17$).
This disparity arises because, in contrast to the budget dilution $\varepsilon/q$ of Local DP, the influence of one individual in Central DP is suppressed to $O(1/n)$, a pattern consistent with the $\sqrt{n}$ gap of Duchi et al.\ \cite{duchi2013local}. However, this utility gap stems from differences in trust models.
The central DP assumes trust in the curator, whereas the Local DP provides stronger privacy guarantees. These two complementary approaches were selected according to the data governance framework.

\subsection{Implications for Biomedical Informatics Practice}

The results translate into several concrete implications for health-informatics workflows.
First, for \textbf{registry-based prognostic research} and \textbf{electronic health record (EHR)-derived survival models}, analysts cannot assume that a fixed privacy budget delivers a single level of utility: predictive discrimination (relevant to risk-stratification tools) and inferential significance (relevant to effect-size reporting and clinical interpretation) recover at different $\varepsilon$ ranges, and $\varepsilon$ chosen for one purpose may be insufficient for the other.
Second, for \textbf{multi-site data collaboration} and \textbf{federated clinical analytics}, the trust-model choice is substantive rather than cosmetic: the Central-DP path (trusted curator releasing perturbed statistics) consistently required less $\varepsilon$ to preserve both predictive and inferential utility than the Local-DP path (each site/participant perturbing raw records), in line with the theoretical $\sqrt{n}$ gap \cite{duchi2013local} and with federated survival frameworks such as dsSurvival \cite{banerjee2022dssurvival}.
Third, for \textbf{privacy-aware model deployment}, our threshold table (Table~\ref{tab:threshold}) provides a dataset-size-aware reference against which governance committees can benchmark proposed $\varepsilon$ settings before a Cox model is released.
Finally, because our bounds are data-driven, these thresholds are best used as optimistic references: actual deployments under formal $\varepsilon$-DP will need either larger $\varepsilon$ or auxiliary techniques (smooth sensitivity \cite{nissim2007smooth}, propose-test-release, or domain-prescribed bounds) to recover comparable utility.

\subsection{Practical Recommendations}

Based on the results of this study, the following guidelines are proposed:
\begin{enumerate}
  \item \textbf{Approach}: Phase~1 (covariate-only perturbation)
    should be the first choice.
    Perturbation of $T/\delta$ should be avoided.
  \item \textbf{$\varepsilon$ setting}
    (exploratory $= \Delta C \leq 0.05$,
    confirmatory $=$ LSR $\leq 0.50$):
    Small datasets ($n < 500$):
    exploratory $\varepsilon \geq 10$,
    confirmatory $\varepsilon \geq 30$;
    Medium datasets ($n \approx 1{,}000$--$3{,}000$):
    exploratory $\varepsilon \geq 5$,
    confirmatory $\varepsilon \geq 15$;
    Large datasets ($n > 5{,}000$):
    exploratory $\varepsilon \geq 3$,
    confirmatory $\varepsilon \geq 10$.
    However, these depend on $q$ and covariate range;
    refer to Table~\ref{tab:threshold} for specifics.
  \item \textbf{Variable management}: Reducing $q$ directly improves utility, as evident by the slow recovery of pbc ($q = 17$).
  \item \textbf{Bound setting}: Clinically reasonable tight bounds are important.
    Range restriction should be considered
    for variables with large ranges.
  \item \textbf{Equal budget allocation}: As discussed in Section~III-A,
    this is a justified design choice
    considering the privacy cost of adaptive allocation.
  \item \textbf{Caution in significance determination}:
    When determining significance following DP application,
    the possibility of false positives should be considered
    for variables whose baseline $p$-values are near
    the $\alpha$ boundary (Section~IV-D).
\end{enumerate}

\subsection{Methodological Limitations}

\begin{enumerate}
  \item \textbf{Data-driven clipping bounds}:
    Formal DP guarantees do not hold.
    Since formal DP requires data-independent bounds
    and noise increases accordingly,
    the results of this study represent an optimistic lower bound.
  \item \textbf{Sensitivity estimation for output perturbation}:
    The dfbeta sensitivity is data-driven
    and not a formal worst-case bound.
    Re-evaluation using data-adaptive sensitivity methods,
    such as smooth sensitivity \cite{nissim2007smooth},
    and alternative perturbation strategies,
    such as objective perturbation,
    \cite{chaudhuri2011differentially} remains warranted.
  \item \textbf{Comparison with prior work}:
    A direct comparison with the output perturbation method
    of Nguyen \& Hui \cite{nguyen2017differentially}
    was not conducted.
    The novelty of Phase~3 is limited to
    its evaluation as an input perturbation variant.
  \item \textbf{Perturbation scope of Phase~1}:
    Phase~1 perturbs only covariates
    whereas $T/\delta$ remain unperturbed;
    therefore, it constitutes covariate-level Local DP,
    not full Local DP.
    The utility of Phase~1 represents
    the optimistic upper bound of input perturbation.
  \item \textbf{Limitation of five datasets}:
    Data with $n > 10{,}000$ have not been verified,
    and extrapolation to large scales
    depends on theoretical predictions.
\end{enumerate}

\subsection{Future Work}

Future research should focus on developing privacy protection methods that preserve statistical utility.
As demonstrated here, the DP-style perturbation schemes evaluated under data-driven bounds essentially eliminate Cox-regression inference at standard $\varepsilon$ levels, with practical recovery requiring $\varepsilon \geq 10$--30 for the settings we studied.
Reconciling privacy protection and utility in medical research requires the following: (1) the development of methods that minimize utility degradation while maintaining formal DP guarantees; (2)~extension to federated learning settings \cite{rahimian2022practical,hung2025federated} to avoid sharing raw data; and (3) an empirical demonstration of scaling effects in large-scale data with $n > 10{,}000$.

\section{Conclusion}

This study systematically and quantitatively evaluates the impact of applying DP mechanisms under data-driven bounds on the statistical utility of the Cox proportional hazards model across 5 clinical datasets ($n = 168$--$6{,}524$) and 15 levels of $\varepsilon$.
Data-driven bounds were used, and the results constituted an optimistic lower bound under the formal DP.
The five key findings are as follows: (1)~\textbf{$\varepsilon \leq 1$ eliminates meaningful Cox inference across all evaluated scales}---LSR $> 0.89$ and C-index $\approx 0.5$ even for the largest dataset; (2)~\textbf{Preservation of risk-set structure is essential}
---perturbation of $T/\delta$ irreversibly impairs discriminative ability; (3)~\textbf{The effect of $n$ first manifests at $\varepsilon \geq 3$}---scaling effects in the transition zone $\varepsilon = 3$--10; (4)~\textbf{$\varepsilon \geq 10$ is the practical minimum}--- $\varepsilon \geq 10$ for predictive performance and $\varepsilon \geq 30$--60 for significance preservation are required; (5)~\textbf{False positive risk increases at moderate-to-high $\varepsilon$}---FPR increases notably for variables whose baseline $p$-values are near $\alpha$.
These findings provide quantitative decision criteria for medical studies when considering the application of DP.

\section*{Ethical Statement}

All datasets used in this study were previously published, publicly available, and included in the R survival package, with no risk of individual re-identification.
Therefore, an ethics committee review was deemed unnecessary.

\section*{Conflict of Interest}

The authors declare no conflict of interest.

\section*{Data and Code Availability}

All the datasets used in this study are publicly available in the R survival package \cite{therneau2024survival}.
The simulation code, reproducibility pipeline, and tutorial for applying DP to the survival analysis are available on GitHub (\url{https://github.com/fk506cni/dp-surv-util-res}).

\section*{Acknowledgment}

Claude Code (Anthropic, Opus 4.6) and Gemini (Google, 3.1 Pro) were used for code development and English-language editing.

\bibliographystyle{IEEEtran}
\bibliography{refs}

\end{document}